\documentclass[journal,draft,onecolumn,12pt]{IEEEtran}
\usepackage{amsfonts,color,morefloats}
\usepackage{amssymb,amsmath,latexsym,amsthm}
\usepackage{float}
\usepackage{algorithm}
\usepackage{algorithmic}
\usepackage{multirow,tabu}
\usepackage{nicematrix} 
\usepackage{amsmath}    
\usepackage{multirow, bigdelim} 
\usepackage{enumitem}   
\usepackage{xcolor}

\newcommand{\PRM}{{\mathrm{PRM}}}

\newcommand{\C}{{\mathcal{C}}}
\newcommand{\D}{{\mathcal{D}}}

\newtheorem{theorem}{Theorem}
\newtheorem{lemma}[theorem]{Lemma}

\newtheorem{example}[theorem]{Example}

\begin{document}

\title{A new class of optimal binary linear complementary pairs of codes}

\author{Yansheng Wu,\thanks{Y. Wu is with the School of Computer Science, Nanjing University of Posts and Telecommunications, Nanjing, 210023, China,  and the State Key Laboratory of Cyberspace Security Defense (Institute of Information Engineering, Chinese Academy of Sciences, Beijing 100085), China. 
 Email: yanshengwu@njupt.edu.cn.} 
\and Jing Chen\thanks{J. Chen is with the School of Computer Science, Nanjing University of Posts and Telecommunications, Nanjing
210023, China. Email: 1806108644@qq.com.} 
}

\date{\today}
\maketitle

\begin{abstract} 
{The construction of optimal binary linear complementary pair (LCP) has recently attracted considerable attention in coding theory.}  Based on the Belov codes, this paper proposes a novel construction of optimal binary LCPs with $\theta=4$, extending our previous work on $\theta=3$. Our results not only yield abundant new optimal binary LCPs with flexible parameters, but also enable more optimal binary LCPs with smaller minimum distances. 
\end{abstract}


\section{Introduction}\label{sec-intro} 
A pair of linear codes $(\C, \D)$ of length $n$ over $\mathbb{F}_{q}$ is called a linear complementary pair (LCP) if their direct sum $\C \oplus \D$ yields the full space $\mathbb{F}_{q}^{n}$. {This is equivalent to $\dim \C + \dim \D = n$ and $\C \cap \D = \{\mathbf{0}\}$.} In particular, if $\dim \C = k$ and $\dim \D = n - k$, {then the $(\C,\D)$ is referred to as an $[n, k]$ LCP.} Clearly, the LCP serve as a natural generalization of linear complementary dual codes (LCD codes), see \cite{c18}. The most important problem in this research is the construction of LCPs achieving the maximal security parameter. For given $n$ and $k$, let
\[
d_{LCP}(n,k) = \max \left\{ \min \left\{ d(\C), d( \D^{\bot}) \right\} : (\C,\D) \text{ is an } [n,k] \text{ binary LCP} \right\}
\]
to be the maximum security parameter of  $[n,k]$ binary LCP. Let $d_L(n,k)$ denote the highest minimum distance among binary linear $[n,k]$ codes, it follows that 
$
 d_{LCP}(n,k) \leq d_L(n,k).
$
We call an $[n,k]$ linear code optimal if its minimum distance is $d_L(n,k)$. Similarly, an $[n,k]$ LCP $(\C, \D)$ is said to be optimal whenever its security parameter reaches $d_L(n,k)$. LCPs have been used in the framework of direct sum masking, which was proposed as a countermeasure against side channel attacks (SCAs) and fault injection attacks (FIAs), see \cite{Carlet18, G23, Lixia}.

Choi {\it et al.} \cite{Choi23} systematically studied the optimality of binary LCPs, derived a sufficient condition for non-optimal binary LCPs, computed all security parameters for lengths up to 18, and constructed infinite families of optimal binary LCPs for $k \ge 2$ and $n \equiv 0,1 \pmod{2^k-1}$ as well as all lengths for $2\le k \le4$ with only two exceptions.
{Li {\it et al.} \cite{Li25} completely solved an open problem on binary LCPs proposed by Carlet {\it et al.}  \cite{Carlet18}, and proved a conjecture on the security parameters of binary LCPs.}
Güneri \cite{G23} constructed a new infinite family of optimal binary LCPs for $k \ge 5$ and large minimum distances based on Solomon–Stiffler codes that achieve the Griesmer bound. Most recently, Wu and Chen \cite{WC} extend the construction of LCPs to Belov codes.

{As we know, Belov codes are a  generalized class of Solomon-Stiffler codes.} In this paper, based on previous work on $\theta=3$, we further delete columns from $\theta=4$ to obtain more binary optimal LCPs.

\section{Preliminaries} 

In this section, we will recall some need results. 

\subsection{Griesmer bound}

Let $q$ be a prime power  and $\mathbb{F}_{q}$ be the finite field with $q$ elements. Let $\mathbb{F}_{q}^{n}$ be an $n$-dimensional vector space over $\mathbb{F}_{q}$. Let $\mathcal{C}$ be a linear code over $\mathbb{F}_q$ with parameters $[n,k,d]_q$, where $n$ is the length of the code, $k$ is its dimension and $d$ is the minimum distance. In coding theory, constructing optimal codes is essential for improving the reliability of communication systems. A linear code $\mathcal{C}$ with parameters $[n,k,d]_q$ is called optimal if its minimum distance $d$ meets a known theoretical bound. See \cite{HP03} for the detail. 

\begin{lemma}{ \cite{G}  \label{Griesmer}
(Griesmer bound)  Let $\mathcal{C}$ be an $[n,k,d]_{q}$-code. Then
$$n \geq g_q(k,d)= \sum\limits_{i=0}^{k-1}\bigg\lceil\frac{d}{q^i}\bigg\rceil,$$  
where $\lceil{\cdot}\rceil$ denotes the ceiling function.  }
\end{lemma}

\subsection{Solomon-Stiffler codes}

In this subsection, we will recall binary Solomon-Stiffler (SS) codes.

Let ${G} = [S_{k}]$ be a $k \times (2^{k} - 1)$ matrix whose columns comprise all nonzero vectors in $\mathbb{F}_{2}^{k}$. Then ${G}$ generates the $[2^{k} - 1, k, 2^{k - 1}]$ simplex code where all the codewords have weight $2^{k - 1}$. We denote the matrix
\[
 \overbrace{  \left[ S_k | \cdots | S_k  \right] }^{h ~\text{times}}  
\]
by $S_{k}(h)$. The code generated by this matrix, which we will denote by $\mathcal C_{{S}_{k}(h)}$, is the $h$-times replicated simplex code. It is clear that $\mathcal C_{{S}_{k}(h)}$ is also a constant-weight code with parameters $[h(2^{k} - 1), k, h2^{k - 1}]$.

Let $\displaystyle F = \bigcup_{i = 1}^{t} U_i$, where $1 \leq u_1 <  \cdots < u_t < k$. Let $U_1, \ldots, U_t$ be subspaces of $\mathbb{F}_2^k$ with $\dim U_i = u_i$, for $i = 1, \ldots, t$. We denote by $U_i$ the $k \times (2^{u_i} - 1)$ matrix of rank $u_i$ whose columns comprise the nonzero elements of the space $U_i$. Clearly, $U_i$ is contained in $S_k$ for all $i$, and the nonzero rows of such a matrix have weight $2^{u_i - 1}$. 
Provided that $U_1, \ldots, U_t$ are chosen to satisfy
\begin{equation}\label{eq1}
\sum_{i = 1}^{t} u_i \leq hk 
\end{equation}
distinct basis elements can be selected for all $U_i$ from the columns of $S_k(h)$, which contains $hk$ basis elements in its columns. Consequently, $F$ can be formed by the columns derived from the nonzero elements of the $U_i$, as condition in \eqref{eq1} ensures that no column in $S_k(h)$ is punctured more than $h$ times. 

\begin{lemma}\label{lem2} (\cite{B74, S65}) 
If $1 \leq u_1 < u_2 < \cdots < u_t < k$
 with $\sum_{i=1}^{t}u_i\le hk$, then ${G} = [S_{k}(h) \setminus F]$ generates  a binary  SS code with parameters
\begin{equation}\label{eqss}
    [n,k,d]=[ h(2^k - 1) - \sum_{i = 1}^{t} (2^{u_i} - 1),k,
 h2^{k - 1} - \sum_{i = 1}^{t} 2^{u_i - 1}].
\end{equation}
This code meets the Griesmer bound.
\end{lemma}

\subsection{Belov codes}
In this subsection, we will recall binary Belov codes, which is a generalization of SS codes. SS codes and Belov codes are two well-known Griesmer codes, see \cite{T92}.

Let $F_1 = W \setminus A$ where $W$ is a subspace of $\mathbb{F}_2^k$ with $\dim W = \theta \geq 3$, $A \subset W$, $\theta - 1 < |A| \leq \theta + 1$ and any set of $\leq \theta$ points in $A$ are linearly independent. Belov codes additionally remove $F_1 = W \setminus A$, yielding a larger family of parameters that covers cases unreachable by SS codes.

\begin{lemma}\label{lem3}(\cite{B74, T92})
If $3\le \theta\le u_1<\cdots< u_t<k$
 with $\sum_{i=1}^{{min}\{h+1,t\}}u_{t+1-i}\le hk$ and $F \cap F_1 = \emptyset$, then ${G} = [S_{k}(h) \setminus (F \cup F_1)]$ generates a binary Belov code with parameters
\begin{equation}
    [n,k,d]=[h(2^k - 1) - \sum_{i = 1}^{t} (2^{u_i} - 1) - \sum_{i = 1}^{\theta - 2} ( 2^{i + 1} - 1 )- \varepsilon,k,h2^{k - 1} - \sum_{i = 1}^{t} 2^{u_i - 1} - \sum_{i = 1}^{\theta - 2} 2^{i} - \varepsilon]
\end{equation}
where $0 \leq \varepsilon = \theta + 1 - |A| \leq 1$.
This code meets the Griesmer bound.
\end{lemma}

\begin{lemma}(\cite[Theorem 3]{B742})\label{lem4} For any two integers $ k$ and $d = h2^{k - 1} - \sum_{i = 1}^{t} 2^{u_i - 1}$ with $1 \leq u_1 < u_2 < \cdots < u_t < k$. Then there exists a $[g_2(k,d), k,d]$ of the Solomon-Stiffler type or of the Belov type if either 
 $$\sum_{i=1}^{{min}\{h+1,t\}}u_{t+1-i}\le hk $$ or
 $$u_{i+1} - u_{i} = 1, ~ \mbox{ for } i=1, 2, \cdots,t-1, ~\mbox{ and }u_1 \in \{1,2\}.$$
    
\end{lemma}

{
\subsection{Linear complementary pair (LCP)}
In this section, we introduce the LCP discriminant criterion, which serves as the key tool for verifying LCPs in subsequent analysis.

\begin{lemma}(\cite[Proposition 13]{Choi23})
     Let $G=(I_k : M )$ and $H=(N:I_{n-k})$ be generator matrices of length n binary linear codes $\C$ and $\D$, respectively, where I denotes the identity matrix of the specified size. Then $(\C , \D)$ is LCP if and only if $I_k+MN$ (or $I_{n-k}$ + NM) is invertible.
\end{lemma}

The construction strategy for optimal binary linear complementary pairs is stated below:
\begin{enumerate}
    \item Choose an optimal binary linear $[n,k]$ code $\C$ whose generator matrix takes the partitioned form $G=(I_k : M )$.
    \item Construct another $k \times n$ matrix $\tilde{H} = (I_k:N^T)$ satisfying the following two conditions: 
    \begin{enumerate}
        \item The code $\D^T$ generated by $\tilde{H}$ is optimal and shares the same minimum distance with $\C$.
        \item $I_k+MN$ is an invertible matrix.
    \end{enumerate}
\end{enumerate} 

\begin{lemma}(\cite[Theorem 4]{G23})
    For any $k \ge 5$ and $d \ge \lceil (k-1)/2\rceil 2^{k-1}$, there exists an optimal binary $[g(k,d), k]$ LCP of codes $(\C,\D)$ with the security parameter $d$, in which $\C$ is a Solomon–Stiffler code.
\end{lemma}
}

\section{More new optimal binary LCPs with $\theta=4$}

In this section, we will obtain more new optimal binary LCPs in case of $\theta=4$.

Before that, the following lemma is very important.
\begin{lemma}\label{lem12}
    Assume that $k \ge 7$ and $d = h2^{k - 1} - \sum_{i = 1}^{t} 2^{u_i - 1}$ with  $4 \leq u_1 < u_2 < \cdots < u_t < k$ and $\sum_{i=1}^{{min}\{h+1,t\}}u_{t+1-i}\le hk $.
    Let $F = \cup_{i = 1}^{t} U_i   $ with $\mbox{dim}(U_i)=u_i$ and $F_1 = W \setminus A$ for some subspace $W$ such that $F\cap F_1=\emptyset$. By choosing appropriately, $G = [S_k(h) \setminus (F \cup F_1)]$ of \(\C\) has the following properties:
    
    \begin{enumerate}
    \item If $h=1$, i.e. $d \le 2^{k-1}$, {it is required that $u_1=k-3\ge4$, which implies $k \ge 7 $,} then $G$ has

    \begin{enumerate}
        \item the $k \times k$ identity matrix $I_k$,
        \item the following column vectors: \\
        $x_1 = (0,1,0,1,1,1,1,\ldots,1)^T$, 
        $x_2 = (1,0,1,0,0,1,1,\ldots,1)^T$, \\
        $x_3 = (1,1,0,0,\ldots,0,1,0)^T$, 
        $x_4 = (1,1,1,0,\ldots,0,0,1)^T$.         
    \end{enumerate}
    \item If $2 \le h \le \lceil \frac{k-1}{2} \rceil$, i.e. $ 2^{k-1} < d \le \lceil \frac{k-1}{2} \rceil 2^{k - 1}$ and  { $k \ge 5$ }, then $G$ has
    \begin{enumerate}
        \item at least one copy of the  $k \times k$ identity matrix $I_k$,
        \item at least one copy of the following column vectors:\\
        $x_5 = (1,1,1,1,1,\ldots,1)^T$,
        $x_6 = (0,0,0,0,1,\ldots,1)^T$,\\
        $x_7 = (1,1,0,\ldots,0,0,1,1)^T$, 
        $x_8 = (0,1,0,\ldots,0,1,0,0)^T$, \\
        $x_9 = (1,1,\ldots,1)^T$, 
        $x_{10} = (0,1,\ldots,1)^T$.\\
        $x_{11} =(0,1,1,0,\cdots,0,1,1,1,1,1,0)^T$,
        $x_{12} =(0,0,0,0,\cdots,0,1,0,0,0,0,1)^T$,\\
        $x_{13}=(1,\cdots,1,0,1,1,0,1,1,1,0,1)^T$.
    \end{enumerate}
    
\end{enumerate}
\end{lemma}

\begin{proof} We first construct the Solomon–Stiffler code, then obtain the Belov code by further deleting $F_1$ from the SS codes.

    \begin{enumerate}
        \item ($h=1$, $k \ge 7$)           
        \begin{enumerate}
            \item (SS code)  
            $u_t\neq k-1$ or $k-2$, to keep the identity $I_k$ in $G$, we have to delete the subspaces of $U_{k-4}$ from \cite[(9)]{G23}. Consider the maximal anticode case, we can delete  the space spanned by the set $\mathcal{B} \setminus (\{ \left( {0,1,0,\ldots ,0,1}\right) \} \cup \{ \left( {0,0,1,0,\ldots ,0,1}\right) \})$. This deletion of maximal subspace ensures the existence of $I_k$, $x_1$, $x_2$, $x_3$ and $x_4$.
            \item (Belov code) Choose $\theta = 4$, $\text{dim}(W)=\theta=4$.

            \begin{enumerate}
                \item $\varepsilon = \theta +1-|A| =0$ for $|A|=\theta+1$, then we choose a basis for $W$: 
                \[
                    w_1 = \begin{pmatrix}1\\0\\0\\1\\0\\1\\0\\ \vdots \\ 0\end{pmatrix},\quad
                    w_2 = \begin{pmatrix}0\\1\\0\\1\\0\\1\\0\\ \vdots\\ 0\end{pmatrix},\quad
                    w_3 = \begin{pmatrix}0\\0\\1\\0\\0\\1\\0\\ \vdots\\0 \end{pmatrix},\quad
                    w_4 = \begin{pmatrix}0\\0\\0\\0\\1\\0\\0\\ \vdots\\0 \end{pmatrix}
                    \hspace{-0.4em}
                    \raisebox{-5.2em}{$\left.\rule{0pt}{2.5em}\right\}$}
                    \hspace{0.3em}
                    \raisebox{-5.2em}{$\text{all-zero } (k-6)\text{ rows,}$}
                    \] and choose $A=\{ w_4, w_4+w_3, w_4+w_2, w_4+w_1, w_1+w_2+w_3 \}$, then $F_1$ can be written as shown below:
                    \[
                    F_1 =
                    \left(
                    \begin{matrix} 
                    0 & 0 & 0 & 0 & 1 & 1 & 1 & 1 & 1 & 1\\
                    0 & 1 & 1 & 1 & 0 & 0 & 0 & 1 & 1 & 1\\
                    1 & 0 & 1 & 1 & 0 & 1 & 1 & 0 & 0 & 1\\
                    0 & 1 & 1 & 1 & 1 & 1 & 1 & 0 & 0 & 0\\
                    0 & 0 & 0 & 1 & 0 & 0 & 1 & 0 & 1 & 1\\
                    1 & 1 & 0 & 0 & 1 & 0 & 0 & 0 & 0 & 1\\
                    0 & 0 & 0 & 0 & 0 & 0 & 0 & 0 & 0 & 0\\
                    \vdots & \vdots & \vdots & \vdots & \vdots & \vdots & \vdots & \vdots & \vdots & \vdots\\
                    0 & 0 & 0 & 0 &  0 & 0 & 0 & 0 & 0 & 0\\
                    \end{matrix}
                    \right)
                    \hspace{-0.4em}
                    \raisebox{-5.3em}{$\left.\rule{0pt}{2.5em}\right\}$}
                    \hspace{0.3em}
                    \raisebox{-5.3em}{$\text{all-zero } (k-6)\text{ rows,}$}
                    \]
                    ensures the existence of $x_1$ , $x_2$.
                
                \item $\varepsilon = \theta +1-|A| =1$ for $|A|=\theta$, 
                 then we choose a basis for $W$: 
                    \[
                    w_1 = \begin{pmatrix}1\\0\\0\\0\\0\\ \vdots \\0 \\ 0\end{pmatrix},\quad
                    w_2 = \begin{pmatrix}0\\1\\0\\0\\0\\ \vdots\\0 \\ 0\end{pmatrix},\quad
                    w_3 = \begin{pmatrix}0\\0\\1\\0\\ \vdots\\0 \\1\\ 1\end{pmatrix},\quad
                    w_4 = \begin{pmatrix}0\\0\\0\\1\\1\\0\\  \vdots\\0 
                    \end{pmatrix}
                    \hspace{-0.4em}
                    \raisebox{-4.4em}{$\left.\rule{0pt}{2.5em}\right\}$}
                    \hspace{0.3em}
                    \raisebox{-4.4em}{$\text{all-zero } (k-5)\text{ rows,}$}
                    \] and choose $A=\{ w_1, w_2, w_3, w_4\}$
                    $F_1$ can be written as shown below:
                    \[
                    F_1 = 
                    \left(
                    \begin{array}{*{11}{c}}
                    0 & 0 & 0 & 0 & 1 & 1 & 1 & 1 & 1 & 1 & 1\\
                    0 & 1 & 1 & 1 & 0 & 0 & 0 & 1 & 1 & 1 & 1\\
                    1 & 0 & 1 & 1 & 0 & 1 & 1 & 0 & 0 & 1 & 1\\
                    1 & 1 & 0 & 1 & 1 & 0 & 1 & 0 & 1 & 0 & 1\\
                    1 & 1 & 0 & 1 & 1 & 0 & 1 & 0 & 1 & 0 & 1\\
                    0 & 0 & 0 & 0 & 0 & 0 & 0 & 0 & 0 & 0 & 0\\
                    \vdots & \vdots & \vdots & \vdots & \vdots & \vdots & \vdots & \vdots & \vdots & \vdots & \vdots \\
                    0 & 0 & 0 & 0 & 0 & 0 & 0 & 0 & 0 & 0 & 0 \\
                    1 & 0 & 1 & 1 & 0 & 1 & 1 & 0 & 0 & 1 & 1 \\
                    1 & 0 & 1 & 1 & 0 & 1 & 1 & 0 & 0 & 1 & 1 \\
                    \end{array}
                    \right)
                    \hspace{-0.4em}
                    \raisebox{-2.5em}{$\left.\rule{0pt}{2.5em}\right\}$}
                    \hspace{0.3em}
                    \raisebox{-2.5em}{$\text{all-zero } (k-7)\text{ rows,}$}
                    \]
                    ensures the existence of $x_3$ , $x_4$.
            \end{enumerate}
        \end{enumerate}
        
        \item ($2 \le h \le \lceil \frac{k-1}{2} \rceil$, { $k \ge 5$})
            \begin{enumerate}
                \item (SS code)
                Let us consider the \lq\lq maximal anticode\rq\rq  case for $u_1 \geq 4$ by deleting subspaces $U_1,  \dots, U_{k-4} \subset {\mathbb{F}}_{2}^{k} $ with dimensions 
                \[
                u_1 = 4, u_2=5, \dots, u_{k-5}=k-2, u_{k-4}=k-1
                \]
                and $\sum_{i=1}^{{min}\{h+1,t\}}u_{t+1-i}\le hk $.
                

                \item (Belov code) To satisfy the Griesmer code, only one copy of $F_1$ can be deleted from the simplex code, and we choose to perform the deletion in $U$ of dimension $k-3$ for the maximal anticode case.

                \begin{enumerate}
                    \item $t<h$, $k \ge 5$
                        \begin{enumerate}
                            \item $\varepsilon = \theta +1-|A| =0$ for $|A|=\theta+1$, We choose a basis for $W$ as
                             \[
                            w_1 = \begin{pmatrix}1\\0\\ 0\\ 0 \\ 0\\ \vdots \\0\end{pmatrix},\quad
                            w_2 = \begin{pmatrix}0\\1 \\ 0\\ 0 \\ 0\\  \vdots \\0\end{pmatrix},\quad
                            w_3 = \begin{pmatrix}0 \\0\\1\\ 0 \\ 0\\ \vdots\\0\end{pmatrix},\quad
                            w_4 = \begin{pmatrix}0 \\0\\ 0\\1 \\ 0\\ \vdots \\ 0 \end{pmatrix}
                            \hspace{-0.4em}
                            \raisebox{-3.5em}{$\left.\rule{0pt}{2.5em}\right\}$}
                            \hspace{0.3em}
                            \raisebox{-3.5em}{$\text{all-zero } (k - 4) \text{ rows,}$}.
                            \] Then 
                            $F_1$ can be written as shown below:
                            \[
                            F_1 = 
                            \begin{pmatrix}
                            0 & 0 & 0 & 0 & 1 & 1 & 1 & 1 & 1 & 1\\
                            0 & 1 & 1 & 1 & 0 & 0 & 0 & 1 & 1 & 1 \\
                            1 & 0 & 1 & 1 & 0 & 1 & 1 & 0 & 0 & 1 \\
                            1 & 1 & 0 & 1 & 1 & 0 & 1 & 0 & 1 & 0\\
                            0 & 0 & 0 & 0 & 0 & 0 & 0 & 0 & 0 & 0 \\
                            \vdots & \vdots & \vdots  & \vdots & \vdots & \vdots  & \vdots & \vdots & \vdots  & \vdots\\
                            0 & 0 & 0 & 0 & 0 & 0 & 0 & 0 & 0 & 0 
                            \end{pmatrix}
                            \hspace{-0.4em}
                            \raisebox{-3.5em}{$\left.\rule{0pt}{2.5em}\right\}$}
                            \hspace{0.3em}
                            \raisebox{-3.5em}{$\text{all-zero } (k - 4) \text{ rows,}$}.
                            \]
                            ensures the existence of $x_5$ , $x_6$.
                            \item $\varepsilon = \theta +1-|A| =1$ for $|A|=\theta$, We choose a basis for $W$ as
                            \[
                            w_1 = \begin{pmatrix}1\\0\\ 0\\ 0 \\ 0\\ \vdots \\0\end{pmatrix},\quad
                            w_2 = \begin{pmatrix}0\\1 \\ 0\\ 0 \\ 0\\  \vdots \\0\end{pmatrix},\quad
                            w_3 = \begin{pmatrix}0 \\0\\1\\ 0 \\ 0\\ \vdots\\0\end{pmatrix},\quad
                            w_4 = \begin{pmatrix}0 \\0\\ 0\\1 \\ 0\\ \vdots \\ 0 \end{pmatrix}
                            \hspace{-0.4em}
                            \raisebox{-3.5em}{$\left.\rule{0pt}{2.5em}\right\}$}
                            \hspace{0.3em}
                            \raisebox{-3.5em}{$\text{all-zero } (k - 4) \text{ rows,}$}.
                            \] Then 
                            $F_1$ can be written as shown below:
                            \[
                            \left(
                            \begin{array}{*{11}{c}}
                            0 & 0 & 0 & 0 & 1 & 1 & 1 & 1 & 1 & 1 & 1\\
                            0 & 1 & 1 & 1 & 0 & 0 & 0 & 1 & 1 & 1 & 1\\
                            1 & 0 & 1 & 1 & 0 & 1 & 1 & 0 & 0 & 1 & 1\\
                            1 & 1 & 0 & 1 & 1 & 0 & 1 & 0 & 1 & 0 & 1\\
                            0 & 0 & 0 & 0 & 0 & 0 & 0 & 0 & 0 & 0 & 0\\
                            \vdots & \vdots & \vdots  & \vdots & \vdots & \vdots  & \vdots & \vdots & \vdots  & \vdots& \vdots\\
                            0 & 0 & 0 & 0 & 0 & 0 & 0 & 0 & 0 & 0 & 0
                            \end{array}
                            \right)
                            \hspace{-0.4em}
                            \raisebox{-3.5em}{$\left.\rule{0pt}{2.5em}\right\}$}
                            \hspace{0.3em}
                            \raisebox{-3.5em}{$\text{all-zero } (k - 4) \text{ rows,}$}.
                            \]
                            there is no need to guarantee the existence of any two columns.
                        \end{enumerate}
                    \item $t=h$, $k \ge 7$
                        \begin{enumerate}
                        \item $\varepsilon = \theta +1-|A| =0$ for $|A|=\theta+1$, We choose a basis for $W$ as
                        \[
                        w_1 = \begin{pmatrix}1\\0\\ \vdots\\ 0\\ 0 \\ 0 \\0\end{pmatrix},\quad
                        w_2 = \begin{pmatrix}0\\0 \\  \vdots \\ 0\\ 1\\ 0 \\0\end{pmatrix},\quad
                        w_3 = \begin{pmatrix}0 \\0\\ \vdots \\ 0\\ 0\\ 1 \\0\end{pmatrix},\quad
                        w_4 = \begin{pmatrix}0 \\0\\ \vdots \\ 0\\ 0\\ 0 \\1\end{pmatrix}
                        \hspace{-0.4em}
                        \raisebox{1.7em}{$\left.\rule{0pt}{2.5em}\right\}$}
                        \hspace{0.3em}
                        \raisebox{1.7em}{$\text{all-zero } (k - 4) \text{ rows,}$}.
                        \] Then 
                        $F_1$ can be written as shown below:
                            \[
                            F_1 = 
                            \begin{pmatrix}
                            0 & 0 & 0 & 0 & 1 & 1 & 1 & 1 & 1 & 1\\
                            0 & 0 & 0 & 0 & 0 & 0 & 0 & 0 & 0 & 0 \\
                            \vdots & \vdots & \vdots  & \vdots & \vdots & \vdots  & \vdots & \vdots & \vdots  & \vdots\\
                            0 & 0 & 0 & 0 & 0 & 0 & 0 & 0 & 0 & 0 \\
                            0 & 1 & 1 & 1 & 0 & 0 & 0 & 1 & 1 & 1 \\
                            1 & 0 & 1 & 1 & 0 & 1 & 1 & 0 & 0 & 1 \\
                            1 & 1 & 0 & 1 & 1 & 0 & 1 & 0 & 1 & 0\\
                            \end{pmatrix}
                            \hspace{-0.4em}
                            \raisebox{1.7em}{$\left.\rule{0pt}{2.5em}\right\}$}
                            \hspace{0.3em}
                            \raisebox{1.7em}{$\text{all-zero } (k - 4) \text{ rows,}$}.
                            \]
                            ensures the existence of $x_7$ , $x_8$.

                        \item $\varepsilon = \theta +1-|A| =1$ for $|A|=\theta$, 
                        We choose a basis for $W$ as
                        \[
                        w_1 = \begin{pmatrix}0 \\\vdots\\0 \\ 1 \\0 \\0 \\0\end{pmatrix},\quad
                        w_2 = \begin{pmatrix}0 \\\vdots\\0 \\ 0 \\1 \\0 \\0\end{pmatrix},\quad
                        w_3 = \begin{pmatrix}0 \\ \vdots\\0 \\ 0 \\0 \\1 \\0\end{pmatrix},\quad
                        w_4 = \begin{pmatrix}0 \\ \vdots\\0 \\ 0 \\0 \\0 \\1\end{pmatrix}
                        \hspace{-0.4em}
                        \raisebox{3.5em}{$\left.\rule{0pt}{2.5em}\right\}$}
                        \hspace{0.3em}
                        \raisebox{3.5em}{$\text{all-zero } (k - 4) \text{ rows}$,}.
                        \] Then 
                        $F_1$ can be written as shown below:
                            \[
                            F_1 = 
                            \left(
                            \begin{array}{*{11}{c}}
                            0 & 0 & 0 & 0 & 0 & 0 & 0 & 0 & 0 & 0 & 0  \\
                            \vdots & \vdots & \vdots  & \vdots & \vdots & \vdots & \vdots & \vdots & \vdots & \vdots & \vdots\\
                            0 & 0 & 0 & 0 & 0 & 0 & 0 & 0 & 0 & 0 & 0  \\
                            0 & 0 & 0 & 0 & 1 & 1 & 1 & 1 & 1 & 1 & 1 \\
                            0 & 1 & 1 & 1 & 0 & 0 & 0 & 1 & 1 & 1 & 1 \\
                            1 & 0 & 1 & 1 & 0 & 1 & 1 & 0 & 0 & 1 & 1 \\
                            1 & 1 & 0 & 1 & 1 & 0 & 1 & 0 & 1 & 0 & 1 \\
                            \end{array}
                            \right)
                            \hspace{-0.4em}
                            \raisebox{3.5em}{$\left.\rule{0pt}{2.5em}\right\}$}
                            \hspace{0.3em}
                            \raisebox{3.5em}{$\text{all-zero } (k - 4) \text{ rows}$,}.
                            \]
                        ensures the existence of $x_9$ , $x_{10}$.
                
                    \end{enumerate}

               {
                \item $t>h$, $k \ge 9$
                    \begin{enumerate}
                        \item $\varepsilon = \theta +1-|A| =0$ for $|A|=\theta+1$, We choose a basis for $W$ as
                            \[
                        w_1 = \begin{pmatrix} 0 \\1 \\0 \\ 0 \\\vdots\\0 \\0 \\0 \\0 \\0 \\1  \end{pmatrix},\quad
                        w_2 = \begin{pmatrix} 0 \\0 \\1 \\ 0 \\\vdots\\0 \\0 \\0 \\0 \\0 \\0 \end{pmatrix},\quad
                        w_3 = \begin{pmatrix} 0 \\0 \\0 \\ 0 \\ \vdots\\0\\1 \\0 \\1 \\0 \\1 \end{pmatrix},\quad
                        w_4 = \begin{pmatrix} 0 \\0 \\0 \\ 0 \\ \vdots\\0 \\0 \\1 \\0 \\1 \\1 \end{pmatrix}
                        \hspace{-0.4em}
                        \raisebox{1.7em}{$\left.\rule{0pt}{2.5em}\right\}$}
                        \hspace{0.3em}
                        \raisebox{1.7em}{$\text{all-zero } (k - 8) \text{ rows}$,}.
                        \] Then 
                        $F_1$ can be written as shown below:
                            \[
                            F_1 = 
                            \left(
                            \begin{array}{*{11}{c}}
                            0 & 0 & 0 & 0 & 0 & 0 & 0 & 0 & 0 & 0   \\
                            0 & 0 & 0 & 0 & 1 & 1 & 1 & 1 & 1 & 1 \\
                            0 & 1 & 1 & 1 & 0 & 0 & 0 & 1 & 1 & 1  \\
                            0 & 0 & 0 & 0 & 0 & 0 & 0 & 0 & 0 & 0   \\
                            \vdots & \vdots & \vdots  & \vdots & \vdots & \vdots & \vdots & \vdots & \vdots & \vdots \\
                            0 & 0 & 0 & 0 & 0 & 0 & 0 & 0 & 0 & 0  \\
                            1 & 0 & 1 & 1 & 0 & 1 & 1 & 0 & 0 & 1  \\
                            1 & 1 & 0 & 1 & 1 & 0 & 1 & 0 & 1 & 0 \\
                            1 & 0 & 1 & 1 & 0 & 1 & 1 & 0 & 0 & 1 \\
                            1 & 1 & 0 & 1 & 1 & 0 & 1 & 0 & 1 & 0  \\
                            0 & 1 & 1 & 0 & 0 & 0 & 1 & 1 & 0 & 0 
                            \end{array}
                            \right)
                            \hspace{-0.4em}
                            \raisebox{1.7em}{$\left.\rule{0pt}{2.5em}\right\}$}
                            \hspace{0.3em}
                            \raisebox{1.7em}{$\text{all-zero } (k - 8) \text{ rows}$,}.
                            \]
                            ensures the existence of $x_{11}$ , $x_{12}$.

                        \item $\varepsilon = \theta +1-|A| =0$ for $|A|=\theta+1$, We choose a basis for $W$ as
                        \[
                        w_1 = \begin{pmatrix}0 \\\vdots\\0 \\ 1 \\0 \\0 \\1 \\0 \\0 \\1 \\0 \\0 \end{pmatrix},\quad
                        w_2 = \begin{pmatrix}0 \\\vdots\\0 \\ 0 \\1 \\0 \\0 \\0 \\0 \\0 \\0 \\0\end{pmatrix},\quad
                        w_3 = \begin{pmatrix}0 \\ \vdots\\0 \\0 \\0 \\1 \\0 \\0 \\1 \\0 \\0 \\0\end{pmatrix},\quad
                        w_4 = \begin{pmatrix}0 \\ \vdots\\0 \\0 \\0 \\0 \\0 \\0 \\0 \\0 \\1 \\0\end{pmatrix}
                        \hspace{-0.4em}
                        \raisebox{8em}{$\left.\rule{0pt}{2.5em}\right\}$}
                        \hspace{0.3em}
                        \raisebox{8em}{$\text{all-zero } (k - 9) \text{ rows}$,}.
                        \] Then 
                        $F_1$ can be written as shown below:
                            \[
                            F_1 = 
                            \left(
                            \begin{array}{*{11}{c}}
                            0 & 0 & 0 & 0 & 0 & 0 & 0 & 0 & 0 & 0 & 0  \\
                            \vdots & \vdots & \vdots  & \vdots & \vdots & \vdots & \vdots & \vdots & \vdots & \vdots & \vdots\\
                            0 & 0 & 0 & 0 & 0 & 0 & 0 & 0 & 0 & 0 & 0  \\
                            0 & 0 & 0 & 0 & 1 & 1 & 1 & 1 & 1 & 1 & 1  \\
                            0 & 1 & 1 & 1 & 0 & 0 & 0 & 1 & 1 & 1 & 1 \\
                            1 & 0 & 1 & 1 & 0 & 1 & 1 & 0 & 0 & 1 & 1 \\
                            0 & 0 & 0 & 0 & 1 & 1 & 1 & 1 & 1 & 1 & 1 \\
                            0 & 0 & 0 & 0 & 0 & 0 & 0 & 0 & 0 & 0 & 0 \\
                            1 & 0 & 1 & 1 & 0 & 1 & 1 & 0 & 0 & 1 & 1 \\
                            0 & 0 & 0 & 0 & 1 & 1 & 1 & 1 & 1 & 1 & 1 \\
                            1 & 1 & 0 & 1 & 1 & 0 & 1 & 0 & 1 & 0 & 1 \\
                            0 & 0 & 0 & 0 & 0 & 0 & 0 & 0 & 0 & 0 & 0 \\
                            \end{array}
                            \right)
                            \hspace{-0.4em}
                            \raisebox{8em}{$\left.\rule{0pt}{2.5em}\right\}$}
                            \hspace{0.3em}
                            \raisebox{8em}{$\text{all-zero } (k - 9) \text{ rows}$,}.
                            \]
                            ensures the existence of $x_9$ , $x_{13}$.
                    \end{enumerate}
                }

                \end{enumerate}
            \end{enumerate}
    \end{enumerate}

    This completes the proof.
\end{proof}

Now we are ready to present our main result as follows:
\begin{theorem}\label{sun-newthm1}
 Assume that $d = h2^{k-1} - \sum_{i=1}^t 2^{u_i-1}$ with $4 \le u_1 < u_2 < \dots < u_t < k$ and$\sum_{i=1}^{\min\{h+1,t\}} u_{t+1-i} \le hk.$
{
Suppose one of the following holds:
\begin{itemize}
    \item $h=1$ and $k\ge7$;
    \item $2\le h \le \lceil \frac{k-1}{2}\rceil$, $t<h$ and $k\ge5$;
    \item $2\le h \le \lceil \frac{k-1}{2}\rceil$, $t=h$ and $k\ge7$;
    \item $2\le h \le \lceil \frac{k-1}{2}\rceil$, $t>h$ and $k\ge9$.
\end{itemize}
}
When $\theta=4$, there exists a binary $[g_2(k,d),k]$ LCP of Belov type with security parameter $d$.
     
\end{theorem}

\begin{proof}
We first construct a Belov code $\C$ with parameters $[g_2(k,d), k, d]$ that satisfies the Griesmer bound.

According to Lemma \ref{lem4}, such a code exists if we select appropriate parameters $\theta$, $A$, and subspaces $U_i$ that satisfy the required conditions.

Let $G = [S_k(h) \setminus (F \cup F_1)]$ be a generator matrix of $\C$, where $F = \bigcup_{i=1}^t U_i$ denotes the union of subspaces for SS codes, and $F_1 = W \setminus A$ represents the additional puncturing set in the Belov construction. By means of suitable column permutations, $G$ can be transformed into the systematic form $G = (I_k : M')$, where $M'$ is a $k \times (n-k)$ matrix. { Let  $n$ be the length of the code and then 
\[
M' = (m_1, m_2, \ldots, m_{n-k}), \quad \text{where } m_i \in \mathbb{F}_2^{k\times 1} \text{ for } 1 \le i \le n-k. 
\] }

The core of the proof lies in analyzing the structure of $M'M'^T$ and constructing an appropriate matrix $N'$, which is a $(n-k) \times k$ matrix, such that $I_k + M'N'$ is invertible. 

\begin{enumerate}
    \item  $h=1$, $t \ge 7$       
    \begin{enumerate}
        \item (SS code) By Lemma \ref{lem12}, we can consider the maximal anticode case with $t=1$ and $u_1=k-3$. Then take $F_1=\emptyset $ and we have $MM^T=I_k$ from  \cite[Theorem 4]{G23}. Then the result follows directly from \cite[Theorem 4]{G23}.
        
        \item (Belov code) By Lemma \ref{lem12}, choose  $\text{dim}(W)=\theta= 4$. Let $\Delta = \sum_{i \in F_1} m_i m_i^T = F_1F_1^T$. {Then $M'M'^T = MM^T +\Delta = I_k + \Delta.$}
        \begin{enumerate}
            \item $\varepsilon = \theta +1-|A| =0$ for $|A|=\theta+1$, then we have 
            \[
            \Delta = F_1F_1^T =
            \left(
            \begin{matrix}
            0 & 1 & 1 & 1 & 1 & 0 & 0 & \cdots & 0 \\ 
            1 & 0 & 1 & 1 & 1 & 0 & 0 & \cdots & 0 \\  
            1 & 1 & 0 & 0 & 1 & 0 & 0 & \cdots & 0 \\ 
            1 & 1 & 0 & 0 & 0 & 0 & 0 & \cdots & 0 \\ 
            1 & 1 & 1 & 0 & 0 & 1 & 0 & \cdots & 0 \\ 
            0 & 0 & 0 & 0 & 1 & 0 & 0 & \cdots & 0 \\ 
            0 & 0 & 0 & 0 & 0 & 0 & 0 & \cdots & 0 \\ 
            \vdots &\vdots &\vdots & \vdots  & \vdots &\vdots &\vdots &\ddots &\vdots\\
            0 & 0 & 0 & 0 & 0 & 0 & 0 & \cdots & 0 \\ 
            \end{matrix}
            \right)
            \hspace{-0.4em}
            \raisebox{-5.2em}{$\left.\rule{0pt}{2.5em}\right\}$}
            \hspace{0.3em}
            \raisebox{-5.2em}{$\text{all-zero } (k-6)\text{ rows,}$}
            \]
            and 
            \[
            M'M'^T =  MM^T + \Delta =
            \begin{pmatrix}
            1 & 1 & 1 & 1 & 1 & 0  \\ 
            1 & 1 & 1 & 1 & 1 & 0  \\  
            1 & 1 & 1 & 0 & 1 & 0   \\ 
            1 & 1 & 0 & 1 & 0 & 0   \\ 
            1 & 1 & 1 & 0 & 1 & 1  \\ 
            0 & 0 & 0 & 0 & 1 & 1  \\ 
             &  &  &  &  &  & I_{k-6}  \\ 
            \end{pmatrix}.
            \]
            Let $x_1 = (0,1,0,1,1,1,1,\ldots,1)^T$ and $x_2 = (1,0,1,0,0,1,1,\ldots,1)^T$. We denote by \(\sigma_1(M')\) the matrix resulting from swapping \(x_1\) and \(x_2\) in \(M'\) and   
            \[
            \varphi_1 = x_1x_1^T + x_2x_2^T + x_1x_2^T + x_2x_1^T= \begin{pmatrix}
            1 & 1 & 1 & 1 & 1 & 0 & \cdots & 0 \\
            1 & 1 & 1 & 1 & 1 & 0 & \cdots & 0 \\
            1 & 1 & 1 & 1 & 1 & 0 & \cdots & 0 \\
            1 & 1 & 1 & 1 & 1 & 0 & \cdots & 0 \\
            1 & 1 & 1 & 1 & 1 & 0 & \cdots & 0 \\
            0 & 0 & 0 & 0 & 0 & 0 & \cdots & 0 \\
            \vdots & \vdots & \vdots& \vdots& \vdots & \vdots &\ddots &\vdots\\
            0 & 0 & 0 & 0 & 0 & 0 & \cdots & 0 \\
            \end{pmatrix},
            \]
            and we cyclically shift row of $\sigma_1(M')$ to obtain the matrix $N'^T$:
            \[
            N'^T =P_1\sigma_1(M') =\begin{pmatrix}
            \sigma_1(M')_3 \\
            \sigma_1(M')_4 \\
            \vdots \\
            \sigma_1(M')_k \\
            \sigma_1(M')_1 \\
            \sigma_1(M')_2
            \end{pmatrix}, 
            ~
            \mbox{ where~} 
            P_1 = \begin{pmatrix}
            0 & 0 & 1 &\cdots&0\\
            \vdots &\vdots &\vdots&\ddots&\vdots \\
            0 & 0 & 0 &\cdots&1\\
            1 & 0 & 0 &\cdots&0\\
            0 & 1 & 0 &\cdots&0
            \end{pmatrix}.
            \]
            Then we have 
            \[
            M'N' = 
            \begin{pmatrix}
            0 \\
            0 & 0 \\
            0 & 1 & 0 \\
            1 & 0 & 1 & 0 \\
            0 & 1 & 0 & 1 & 0 \\
            0 & 0 & 1 & 1 & 0 & 0 \\
            0 & 0 & 0 & 0 & 1 & \cdots & 0 \\
            \vdots & \vdots & \vdots & \vdots & \vdots & \ddots & \vdots &\ddots\\
            0 & 0 & 0 & 0 & 0 & \cdots & 1 & 0 & 0\\
            \end{pmatrix}.
            \]
            Since \(M'N'\) is a lower triangular matrix, which implies that \(I_k + M'N'\) is invertible.

            \item $\varepsilon = \theta +1-|A| =1$ for $|A|=\theta$, then we have  
            \[
            \Delta = F_1F_1^\mathrm{T} =
            \begin{pmatrix}
            1 & 0 & 0 & 0 & 0 & 0 &\cdots & 0 & 0 & 0\\ 
            0 & 1 & 0 & 0 & 0 & 0 &\cdots & 0 & 0 & 0\\ 
            0 & 0 & 1 & 0 & 0 & 0 &\cdots & 0 & 1 & 1\\
            0 & 0 & 0 & 1 & 1 & 0 &\cdots & 0 & 0 & 0\\
            0 & 0 & 0 & 1 & 1 & 0 &\cdots & 0 & 0 & 0 \\
            0 & 0 & 0 & 0 & 0 & 0 &\cdots & 0 & 0 & 0 \\
            \vdots & \vdots & \vdots& \vdots& \vdots& \vdots& \ddots  & \vdots & \vdots& \vdots\\
            0 & 0 & 0 & 0 & 0 & 0 &\cdots & 0 & 0 & 0 \\
            0 & 0 & 1 & 0 & 0 & 0 &\cdots & 0 & 1 & 1 \\
            0 & 0 & 1 & 0 & 0 & 0 &\cdots & 0 & 1 & 1 
            \end{pmatrix}
            \hspace{-0.4em}
            \raisebox{-2.7em}{$\left.\rule{0pt}{2.5em}\right\}$}
            \hspace{0.3em}
            \raisebox{-2.7em}{$\text{all-zero } (k-7)\text{ rows,}$}
            \]
            and 
            \[
            M'M'^T =  MM^T + \Delta =
            \begin{pmatrix}
            0 & 0 & 0 & 0 & 0 & 0 &\cdots & 0 & 0 & 0\\ 
            0 & 0 & 0 & 0 & 0 & 0 &\cdots & 0 & 0 & 0\\ 
            0 & 0 & 0 & 0 & 0 & 0 &\cdots & 0 & 1 & 1\\
            0 & 0 & 0 & 0 & 1 & 0 &\cdots & 0 & 0 & 0\\
            0 & 0 & 0 & 1 & 0 & 0 &\cdots & 0 & 0 & 0 \\
             &  &  &  &  & 1 &\cdots & 0 &  &  \\
             &  &  &  &  & \vdots& \ddots  & \vdots & & \\
             &  &  &  &  & 0 &\cdots & 1 &  &  \\
            0 & 0 & 1 & 0 & 0 & 0 &\cdots & 0 & 0 & 1 \\
            0 & 0 & 1 & 0 & 0 & 0 &\cdots & 0 & 1 & 0 
            \end{pmatrix}
            \hspace{-0.4em}
            \raisebox{-2.7em}{$\left.\rule{0pt}{2.5em}\right\}$}
            \hspace{0.3em}
            \raisebox{-2.7em}{$I_{k-7}.$}
            \]
            Let \(x_3 = (1, 1, 0, 0, \ldots,0, 1,0)^T\) and \(x_4 = (1, 1,1,0, \ldots, 0,0, 1)^T\).  We denote by \(\sigma_2(M')\) the matrix resulting from swapping \(x_3\) and \(x_4\) in \(M'\) and  
            \[
            \varphi_2 = x_3x_3^T + x_4x_4^T + x_3x_4^T + x_4x_3^T= 
            \begin{pmatrix}
            0 & 0 & 0 & 0 &  \cdots & 0 & 0 & 0\\ 
            0 & 0 & 0 & 0 &  \cdots & 0 & 0 & 0\\ 
            0 & 0 & 1 & 0 &  \cdots & 0 & 1 & 1\\
            0 & 0 & 0 & 0 & \cdots  & 0 & 0 & 0 \\
            \vdots & \vdots & \vdots& \vdots& \ddots  & \vdots& \vdots& \vdots \\
            0 & 0 & 0 & 0 & \cdots  & 0 & 0 & 0 \\
            0 & 0 & 1 & 0 &  \cdots & 0 & 1 & 1\\
            0 & 0 & 1 & 0 &  \cdots & 0 & 1 & 1\\
            \end{pmatrix}
            \hspace{-0.4em}
            \raisebox{-0.8em}{$\left.\rule{0pt}{2.5em}\right\}$}
            \hspace{0.3em}
            \raisebox{-0.8em}{$\text{all-zero } (k-5)\text{ rows.}$}
            \]
            We cyclically shift row of $\sigma_2(M')$ to obtain the matrix $N'^T$:
            \[
            N'^T=P_1\sigma_2(M')= \begin{pmatrix}
            \sigma_2(M')_3 \\
            \sigma_2(M')_4 \\
            \vdots \\
            \sigma_2(M')_k \\
            \sigma_2(M')_1 \\
            \sigma_2(M')_2
            \end{pmatrix},
            ~
            \mbox{ where~} 
            P_1 = \begin{pmatrix}
            0 & 0 & 1 &\cdots&0\\
            \vdots &\vdots &\vdots&\ddots&\vdots \\
            0 & 0 & 0 &\cdots&1\\
            1 & 0 & 0 &\cdots&0\\
            0 & 1 & 0 &\cdots&0
            \end{pmatrix}.
            \]
            Namely, the matrix $M'N'$ is obtained by cyclically permuting the columns of $M'\sigma_2(M')^T$:
            Then we have
            \[
            M'N' = 
            \begin{pmatrix}
            0 \\
            0 & 0 \\
            1 & 0 & 0\\
            0& 0& 1&0 \\
            0& 1& 0& 0&0 \\
            0& 0& 0& 1&\cdots &0\\
            \vdots &\vdots &\vdots & \vdots& \ddots&\vdots& \ddots\\
            0 & 0 & 0&0&\cdots&1 &0 &0 \\
            \end{pmatrix}.
            \]
            It is easy to check that \(I_k + M'N'\) is invertible.

        \end{enumerate}
    \end{enumerate}
        
    \item $2 \le h \le \lceil \frac{k-1}{2} \rceil$
        \begin{enumerate}
            \item (SS code) By Lemma \ref{lem12}, we can consider the maximal anticode case with $u_1\ge3$. $\theta =4 $, $u_1 \ge 4$ Then take $F_1=\emptyset $ and we have $MM^T=I_k$ from  \cite[Theorem 4]{G23}. Then the result follows directly from \cite[Theorem 4]{G23}.

            \item (Belov code)
            \begin{enumerate}
                \item $t<h$, $k \ge 5$
                \begin{enumerate}
                    \item $\varepsilon = \theta +1-|A| =0$ for $|A|=\theta+1$, then we have  
                    \[
                    \Delta = F_1F_1^T =
                    \left(
                    \begin{matrix}
                    0  & 1 & 1 & 1 \\
                    1  & 0 & 1 & 1 \\ 
                    1  & 1 & 0 & 1 \\ 
                    1  & 1 & 1 & 0  \\ 
                      & &  & &0&\cdots&0\\ 
                   & & &   & \vdots& \ddots& \vdots \\
                      &  & &  & 0& \cdots & 0
                    \end{matrix}
                    \right)
                    \hspace{-0.4em}
                    \raisebox{-3.5em}{$\left.\rule{0pt}{2.5em}\right\}$}
                    \hspace{0.3em}
                    \raisebox{-3.5em}{$\text{all-zero } (k-4)\text{ rows,}$}
                    \]
                    and
                     \[
                        M'M'^T =  MM^T + \Delta =
                        \begin{pmatrix}
                            1  & 1 & 1 & 1 \\
                            1  & 1 & 1 & 1 \\ 
                            1  & 1 & 1 & 1 \\ 
                            1  & 1 & 1 & 1  \\ 
                              &  &  &  & 1& \cdots & 0\\ 
                            & & &   & \vdots& \ddots& \vdots \\
                              &  &  &  & 0& \cdots & 1
                        \end{pmatrix}
                        \hspace{-0.4em}
                        \raisebox{-3.5em}{$\left.\rule{0pt}{2.5em}\right\}$}
                        \hspace{0.3em}
                        \raisebox{-3.5em}{$I_{k-4}.$}
                        \]
                    Let $x_5 = (1,1,1,1,1,\ldots,1)^T$ and $x_6 = (0,0,0,0,1,\ldots,1)^T$.  We denote by \(\sigma_3(M')\) the matrix resulting from swapping \(x_5\), \(x_6\) and 
                    \[
                    \varphi_3 = x_5x_5^T + x_6x_6^T + x_5x_6^T + x_6x_5^T= 
                    \begin{pmatrix}
                    1  & 1 & 1 & 1 \\
                    1 & 1 & 1 & 1  \\
                    1 & 1 & 1 & 1 \\
                    1 & 1 & 1 & 1  \\
                      &  &  & & 0 & \cdots & 0\\
                      &&&& \vdots &\ddots &\vdots\\
                      &  &  &  & 0 & \cdots & 0
                    \end{pmatrix},
                    \]
                     and we cyclically shift row of $\sigma_3(M')$ to obtain the matrix $N'^T$:        
                        \[
                        N'^T =P_2\sigma_3(M') =\begin{pmatrix}
                        \sigma_3(M')_2 \\
                        \sigma_3(M')_3 \\
                        \vdots \\
                        \sigma_3(M')_k \\
                        \sigma_3(M')_1 
                        \end{pmatrix}, 
                        ~
                        \mbox{ where~} 
                        P_2 = \begin{pmatrix}
                        0 & 1 & 0 &\cdots&0\\
                        0 & 0 & 1 &\cdots&0\\
                        \vdots &\vdots &\vdots&\ddots&\vdots \\
                        0 & 0 & 0 &\cdots&1\\
                        1 & 0 & 0 &\cdots&0
                        \end{pmatrix},
                        \]
                        then we have 
                        \[
                        M'N' = 
                        \begin{pmatrix}
                        0 \\
                        0 & 0\\
                        0 & 0  &0\\
                        0 & 0  &0 &0 \\
                        0 & 0 & 0 & 1 &0 \\
                        \vdots &\vdots & \vdots &\vdots&\ddots&\ddots \\
                        0 &\cdots& \cdots &0&\cdots&1&0 \\
                        \end{pmatrix}.
                        \hspace{-0.4em}
                        \raisebox{-3.5em}{$\left.\rule{0pt}{2.5em}\right\}$}
                        \hspace{0.3em}
                        \raisebox{-3.5em}{$I_{k-4},$}
                        \]
                        which implies that \(I_k + M'N'\) is invertible.

                    \item $\varepsilon = \theta +1-|A| =1$ for $|A|=\theta$,  then we have  
                    \[
                    \Delta = F_1F_1^T =
                    \left(
                    \begin{matrix}
                    1  & 0 & 0 & 0\\
                    0  & 1 & 0 & 0 \\ 
                    0  & 0 & 1 & 0 \\ 
                    0  & 0 & 0 & 1  \\ 
                      &  & &  & 0& \cdots & 0\\ 
                    &&&   & \vdots& \ddots& \vdots \\
                      &  &  &  & 0& \cdots & 0
                    \end{matrix}
                    \right)
                    \hspace{-0.4em}
                    \raisebox{-3.5em}{$\left.\rule{0pt}{2.5em}\right\}$}
                    \hspace{0.3em}
                    \raisebox{-3.5em}{$\text{all-zero } (k-4)\text{ rows,}$}
                    \]
                    and
                    \[
                     M'M'^T =  MM^T + \Delta =
                    \begin{pmatrix}
                    0  & 0 & 0 & 0 \\
                    0  & 0 & 0 & 0 \\ 
                    0  & 0 & 0 & 0 \\ 
                    0  & 0 & 0 & 0 \\ 
                      &  &  &  & 1& \cdots & 0\\ 
                    & & &   & \vdots& \ddots& \vdots \\
                      &  &  &  & 0& \cdots & 1
                    \end{pmatrix}
                    \hspace{-0.4em}
                    \raisebox{-3.5em}{$\left.\rule{0pt}{2.5em}\right\}$}
                    \hspace{0.3em}
                    \raisebox{-3.5em}{$I_{k-4}.$}
                    \]
                    We cyclically shift row of $M'$ to obtain the matrix $N'^T$:
                    \[
                        N'^T =P_2M' =\begin{pmatrix}
                        M'_2 \\
                        M'_3 \\
                        \vdots \\
                        M'_k \\
                        M'_1 
                        \end{pmatrix}, 
                        ~
                        \mbox{ where~} 
                        P_2 = \begin{pmatrix}
                        0 & 1 & 0 &\cdots&0\\
                        0 & 0 & 1 &\cdots&0\\
                        \vdots &\vdots &\vdots&\ddots&\vdots \\
                        0 & 0 & 0 &\cdots&1\\
                        1 & 0 & 0 &\cdots&0
                        \end{pmatrix},
                    \]
                    then we have 
                        \[
                        M'N' = 
                        \begin{pmatrix}
                        0 \\
                        0 & 0\\
                        0 & 0  &0\\
                        0 & 0  &0 &0 \\
                        0 & 0 & 0 & 1 &0 \\
                        \vdots &\vdots & \vdots &\vdots&\ddots&\ddots \\
                        0 &\cdots& \cdots &0&\cdots&1&0 \\
                        \end{pmatrix}.
                        \hspace{-0.4em}
                        \raisebox{-3.5em}{$\left.\rule{0pt}{2.5em}\right\}$}
                        \hspace{0.3em}
                        \raisebox{-3.5em}{$I_{k-4}.$}
                        \]
                    It is easy to check that \(I_k + M'N'\) is invertible.

                \end{enumerate}

                \item $t=h$, $k \ge 7$
            \begin{enumerate}
            
            \item $\varepsilon = \theta +1-|A| =0$ for $|A|=\theta+1$, then we have 
            \[
            \Delta = F_1F_1^T =
            \left(
            \begin{matrix}
            0 & 0& \cdots & 0 & 1 & 1 & 1 \\ 
            0 & 0& \cdots & 0 & 0 & 0 & 0 \\ 
            \vdots& \vdots&\ddots & \vdots  & \vdots& \vdots& \vdots \\
            0 & 0& \cdots & 0 & 0 & 0 & 0 \\ 
            1 & 0& \cdots & 0 & 0 & 1 & 1 \\ 
            1 & 0& \cdots & 0 & 1 & 0 & 1 \\ 
            1 & 0& \cdots & 0 & 1 & 1 & 0 \\ 
            \end{matrix}
            \right)
            \hspace{-0.4em}
            \raisebox{1.8em}{$\left.\rule{0pt}{2.5em}\right\}$}
            \hspace{0.3em}
            \raisebox{1.8em}{$\text{all-zero } (k-4)\text{ rows,}$}
            \]
            and 
            \[
            M'M'^T =  MM^T + \Delta =
            \begin{pmatrix}
            1 & 0& \cdots & 0 & 1 & 1 & 1 \\ 
             & 1& \cdots & 0 &  &  &  \\ 
            & \vdots&\ddots & \vdots   \\
             & 0& \cdots & 1 &  &  &  \\ 
            1 & 0& \cdots & 0 & 1 & 1 & 1 \\ 
            1 & 0& \cdots & 0 & 1 & 1 & 1 \\ 
            1 & 0& \cdots & 0 & 1 & 1 & 1 \\ 
            \end{pmatrix}
            \hspace{-0.4em}
            \raisebox{1.8em}{$\left.\rule{0pt}{2.5em}\right\}$}
            \hspace{0.3em}
            \raisebox{1.8em}{$I_{k-4}.$}
            \]
            Let $x_7 = (1,1,0,\ldots,0,0,1,1)^T$ and $x_8 = (0,1,0,\ldots,0,1,0,0)^T$.  We denote by \(\sigma_4(M')\) the matrix resulting from swapping \(x_7\), \(x_8\) and   
            \[
            \varphi_4 = x_7x_7^T + x_8x_8^T + x_7x_8^T + x_8x_7^T= 
            \begin{pmatrix}
            1 & 0 & \cdots & 0 & 1 & 1 & 1 \\
            0 & 0 & \cdots & 0 & 0 & 0 & 0\\
            \vdots & \vdots &\ddots &\vdots &\vdots&\vdots&\vdots\\
            0 & 0 & \cdots & 0 & 0 & 0 & 0\\
            1 & 0 & \cdots & 0 & 1 & 1 & 1 \\
            1 & 0 & \cdots & 0 & 1 & 1 & 1 \\
            1 & 0 & \cdots & 0 & 1 & 1 & 1 \\
            \end{pmatrix},
            \]
            and we cyclically shift row of $\sigma_4(M')$ to obtain the matrix $N'^T$:
            \[
            N'^T =P_2\sigma_4(M') =\begin{pmatrix}
            \sigma_4(M')_2 \\
            \sigma_4(M')_3 \\
            \vdots \\
            \sigma_4(M')_k \\
            \sigma_4(M')_1
            \end{pmatrix}, 
            ~
            \mbox{ where~} 
            P_2 = \begin{pmatrix}
            0 & 1 & 0 &\cdots&0\\
            0 & 0 & 1 &\cdots&0\\
            \vdots &\vdots &\vdots&\ddots&\vdots \\
            0 & 0 & 0 &\cdots&1\\
            1 & 0 & 0 &\cdots&0
            \end{pmatrix}.
            \]
            then we have 
            \[
            M'N' = 
            \begin{pmatrix}
            0 \\
            1 & 0\\
            \vdots &\ddots &\ddots\\
            0 &\cdots& 1 &0\\
            0 &\cdots & \cdots &\cdots &0 \\
            0 &\cdots & \cdots &\cdots &\cdots&0 \\
            0 &\cdots & \cdots &\cdots&\cdots&\cdots &0 \\
            \end{pmatrix}
            \hspace{-0.4em}
            \raisebox{1.8em}{$\left.\rule{0pt}{2.5em}\right\}$}
            \hspace{0.3em}
            \raisebox{1.8em}{$I_{k-4}.$}
            \]
            It is easy to check that \(I_k + M'N'\) is invertible.

            \item $\varepsilon = \theta +1-|A| =1$ for $|A|=\theta$,  then we have  
            \[
            \Delta = F_1F_1^T =
            \left(
            \begin{matrix}
            0 & \cdots &0\\
            \vdots & \ddots  & \vdots \\
            0 & \cdots & 0  \\ 
            &&&I_4
            \end{matrix}
            \right)
            \hspace{-0.4em}
            \raisebox{0.8em}{$\left.\rule{0pt}{2.5em}\right\}$}
            \hspace{0.3em}
            \raisebox{0.8em}{$\text{all-zero } (k-4)\text{ rows, }$}
            \]
            and 
            \[
            M'M'^T =  MM^T + \Delta =
            \begin{pmatrix}
            I_{k-4}\\
             &0 & \cdots &0 \\
             & \vdots & \ddots & \vdots \\
             &0 &\cdots & 0
            \end{pmatrix}
            \hspace{-0.4em}
            \raisebox{-0.8em}{$\left.\rule{0pt}{2.5em}\right\}$}
            \hspace{0.3em}
            \raisebox{-0.8em}{$\text{all-zero } 4\text{ rows}$}
            \]
            Let $x_9 = (1,1,\ldots,1)^T$ and $x_{10}= (0,1,\ldots,1)^T$. We denote by \(\sigma_5(M')\) the matrix resulting from swapping \(x_9\) and \(x_{10}\) in \(M'\) and   
            \[
            \varphi_5 = x_9x_9^T + x_{10}x_{10}^T + x_9x_{10}^T + x_{10}x_9^T= 
            \begin{pmatrix}
            1 &0 & \cdots & 0 \\
            0  & 0&\cdots & 0 \\
            \vdots& \vdots &\ddots &\vdots\\
            0  &0& \cdots & 0 \\
            \end{pmatrix}.
            \]
            and we cyclically shift row of $\sigma_5(M')$ to obtain the matrix $N'^T$:
            
            \[
            N'^T =P_2\sigma_5(M') =\begin{pmatrix}
            \sigma_5(M')_2 \\
            \sigma_5(M')_3 \\
            \vdots \\
            \sigma_5(M')_k \\
            \sigma_5(M')_1 
            \end{pmatrix}, 
            ~
            \mbox{ where~} 
            P_2 = \begin{pmatrix}
            0 & 1 & 0 &\cdots&0\\
            0 & 0 & 1 &\cdots&0\\
            \vdots &\vdots &\vdots&\ddots&\vdots \\
            0 & 0 & 0 &\cdots&1\\
            1 & 0 & 0 &\cdots&0
            \end{pmatrix},
            \]
            then we have 
            \[
            M'N' = 
            \begin{pmatrix}
            0 \\
            1 & 0\\
            \vdots &\ddots &\ddots\\
            0 &\cdots& 1 &0\\
            0 &\cdots & \cdots &\cdots &0 \\
            0 &\cdots & \cdots &\cdots &\cdots&0 \\
            0 &\cdots & \cdots &\cdots&\cdots&\cdots &0 \\
            0 &\cdots & \cdots &\cdots&\cdots&\cdots &\cdots&0 \\
            \end{pmatrix}.
            \hspace{-0.4em}
            \raisebox{2.8em}{$\left.\rule{0pt}{2.5em}\right\}$}
            \hspace{0.3em}
            \raisebox{2.8em}{$I_{k-5},$}
            \]
            which implies that \(I_k + M'N'\) is invertible.
        \end{enumerate}

{        
        \item $t>h$, $k \ge 9$
            \begin{enumerate}
                \item $\varepsilon = \theta +1-|A| =0$ for $|A|=\theta+1$, then we have 
                 \[
                \Delta = F_1F_1^T =
                \left(
                \begin{array}{*{11}{c}}
                    0& 0& 0& 0& \cdots&0&0& 0& 0& 0& 0\\
                    0& 0& 1& 0& \cdots&0&1& 1& 1& 1& 0\\
                    0& 1& 0& 0& \cdots&0&1& 1& 1& 1& 1\\
                    0& 0& 0& 0& \cdots&0&0& 0& 0& 0& 0\\
                    \vdots&\vdots&\vdots&\vdots&\ddots&\vdots&\vdots&\vdots&\vdots&\vdots&\vdots\\
                    0& 0& 0& 0& \cdots&0&0& 0& 0& 0& 0\\
                    0& 1& 1& 0& \cdots&0&0& 1& 0& 1& 0\\
                    0& 1& 1& 0& \cdots&0&1& 0& 1& 0& 0\\
                    0& 1& 1& 0& \cdots&0&0& 1& 0& 1& 0\\
                    0& 1& 1& 0& \cdots&0&1& 0& 1& 0& 0\\
                    0& 0& 1& 0& \cdots&0&0& 0& 0& 0& 0\\
                \end{array}
                \right)
                \hspace{-0.4em}
                \raisebox{1.7em}{$\left.\rule{0pt}{2.5em}\right\}$}
                \hspace{0.3em}
                \raisebox{1.7em}{$\text{all-zero } (k-8)\text{ rows,}$}
                \]
                and
                \[
                M'M'^T =  MM^T \oplus \Delta =
                \left(
                \begin{array}{*{11}{c}}
                    1& 0& 0& 0& \cdots&0&0& 0& 0& 0& 0\\
                    0& 1& 1& 0& \cdots&0&1& 1& 1& 1& 0\\
                    0& 1& 1& 0& \cdots&0&1& 1& 1& 1& 1\\
                    & & & 1& \cdots&0&& && & \\
                    &&&\vdots&\ddots&\vdots&&&&&\\
                    & & & 0& \cdots&1&& & & & \\
                    0& 1& 1& 0& \cdots&0&1& 1& 0& 1& 0\\
                    0& 1& 1& 0& \cdots&0&1& 1& 1& 0& 0\\
                    0& 1& 1& 0& \cdots&0&0& 1& 1& 1& 0\\
                    0& 1& 1& 0& \cdots&0&1& 0& 1& 1& 0\\
                    0& 0& 1& 0& \cdots&0&0& 0& 0& 0& 1\\
                \end{array}
                \right)
                \hspace{-0.4em}
                \raisebox{1.7em}{$\left.\rule{0pt}{2.5em}\right\}$}
                \hspace{0.3em}
                \raisebox{1.7em}{$I_{k-8}$.}
            \]
            Let $x_{11} = (0,1,1,0,\cdots,0,1,1,1,1,1,0)^T$ 
            and $x_{12} = (0,0,0,0,\cdots,0,1,0,0,0,0,1)^T$.  We denote by \(\sigma_6(M')\) the matrix resulting from swapping \(x_{11}\), \(x_{12}\) and   
            \[
            \varphi_6 = x_{11}x_{11}^T + x_{12}x_{12}^T + x_{11}x_{12}^T + x_{12}x_{11}^T= 
            \left(
                \begin{array}{*{12}{c}}
                 0& 0& 0& 0 & \cdots & 0 & 0 & 0 & 0 & 0 & 0 & 0\\  
                 0& 1& 1& 0 & \cdots & 0 & 0 & 1 & 1 & 1 & 1 & 1\\  
                 0& 0& 0& 0 & \cdots & 0 & 0 & 1 & 1 & 1 & 1 & 1\\  
                 0& 0& 0& 0 & \cdots & 0 & 0 & 0 & 0 & 0 & 0 & 0\\  
                 \vdots&\vdots&\vdots&\vdots&\ddots&\vdots&\vdots&\vdots&\vdots&\vdots&\vdots&\vdots\\
                 0& 0& 0& 0 & \cdots & 0 & 0 & 0 & 0 & 0 & 0 & 0\\  
                 0& 0& 0& 0 & \cdots & 0 & 0 & 0 & 0 & 0 & 0 & 0\\  
                 0& 1& 1& 0 & \cdots & 0 & 0 & 1 & 1 & 1 & 1 & 1\\  
                 0& 1& 1& 0 & \cdots & 0 & 0 & 1 & 1 & 1 & 1 & 1\\  
                 0& 1& 1& 0 & \cdots & 0 & 0 & 1 & 1 & 1 & 1 & 1\\  
                 0& 1& 1& 0 & \cdots & 0 & 0 & 1 & 1 & 1 & 1 & 1\\  
                 0& 1& 1& 0 & \cdots & 0 & 0 & 1 & 1 & 1 & 1 & 1\\    
                \end{array}
                \right), 
            \]
            and we cyclically shift row of $\sigma_6(M')$ to obtain the matrix $N'^T$:
            \[
            N'^T =P_3\sigma_6(M') =\begin{pmatrix}
            \sigma_6(M')_{k-5} \\
            \sigma_6(M')_{k-4} \\
            \vdots \\
            
            \sigma_6(M')_k \\
            \sigma_6(M')_1 \\
            \sigma_6(M')_2 \\
            \vdots \\
            \sigma_6(M')_{k-6}
            \end{pmatrix}, 
            ~
            \mbox{ where~} 
            P_3 = \begin{pmatrix}
            0 & I_6\\
            I_{k-6} & 0
            \end{pmatrix}.
            \]
            Then we have 
            \[
            M'N' = 
            \left(
            \begin{array}{*{13}{c}}
               0 & 0 & 0 & 0 & 0 & 0 & 1& 0 &0  \\
               0 & 0 & 0 & 0 & 0 & 1 & 0& 0 &0  \\
               0 & 0 & 0 & 0 & 0 & 0 & 0& 0 &0  \\
                &  & &  &  &  & &  & &1 & \cdots & 0 \\
                &  & &  & &  & & & & \vdots & \ddots & \vdots \\
                &  &  &  &  & & & & &0 & \cdots & 1 \\
               1 & 0 & 0 & 0 & 0 & 0 & 0& 0 &0  \\
               0 & 0 & 0 & 1 & 0 & 1 & 0& 0 &0  \\
               0 & 0 & 0 & 0 & 1 & 1 & 0& 0 &0  \\
               0 & 1 & 0 & 0 & 0 & 1 & 0& 0 &0  \\
               0 & 0 & 1 & 0 & 0 & 1 & 0& 0 &0  \\
               0 & 1 & 1 & 1 & 1 & 0 & 0& 1 &0  \\
               
            \end{array}
            \right)
            \hspace{-0.4em}
            \raisebox{2.5em}{$\left.\rule{0pt}{2.5em}\right\}$}
            \hspace{0.3em}
            \raisebox{2.5em}{$I_{k-9},$}
            \]

                \item $\varepsilon = \theta +1-|A| =1$ for $|A|=\theta$,  then we have
                \[
                \Delta = F_1F_1^T =
                \left(
                \begin{array}{*{12}{c}}
                0 & \cdots & 0\\ 
                \vdots & \ddots &\vdots\\ 
                0&\cdots &0  \\
                 & &  & 1 & 0 & 0 & 1 & 0 & 0 & 1 & 0 & 0\\  
                 & &  & 0 & 1 & 0 & 0 & 0 & 0 & 0 & 0 & 0\\  
                 & &  & 0 & 0 & 1 & 0 & 0 & 1 & 0 & 0 & 0\\  
                 & &  & 1 & 0 & 0 & 1 & 0 & 0 & 1 & 0 & 0\\  
                 & &  & 0 & 0 & 0 & 0 & 0 & 0 & 0 & 0 & 0\\  
                 & &  & 0 & 0 & 1 & 0 & 0 & 1 & 0 & 0 & 0\\  
                 & &  & 1 & 0 & 0 & 1 & 0 & 0 & 1 & 0 & 0\\  
                 & &  & 0 & 0 & 0 & 0 & 0 & 0 & 0 & 1 & 0\\  
                 & &  & 0 & 0 & 0 & 0 & 0 & 0 & 0 & 0 & 0\\  
                \end{array}
                \right)
                \hspace{-0.4em}
                \raisebox{8em}{$\left.\rule{0pt}{2.5em}\right\}$}
                \hspace{0.3em}
                \raisebox{8em}{$\text{all-zero } (k-9)\text{ rows,}$}
                \]and
                \[
                M'M'^T =  MM^T \oplus \Delta =
                \left(
                \begin{array}{*{12}{c}}
                1 & \cdots & 0\\ 
                \vdots & \ddots &\vdots\\ 
                0&\cdots &1  \\
                 & &  & 0 & 0 & 0 & 1 & 0 & 0 & 1 & 0 & 0\\  
                 & &  & 0 & 0 & 0 & 0 & 0 & 0 & 0 & 0 & 0\\  
                 & &  & 0 & 0 & 0 & 0 & 0 & 1 & 0 & 0 & 0\\  
                 & &  & 1 & 0 & 0 & 0 & 0 & 0 & 1 & 0 & 0\\  
                 & &  & 0 & 0 & 0 & 0 & 1 & 0 & 0 & 0 & 0\\  
                 & &  & 0 & 0 & 1 & 0 & 0 & 0 & 0 & 0 & 0\\  
                 & &  & 1 & 0 & 0 & 1 & 0 & 0 & 0 & 0 & 0\\  
                 & &  & 0 & 0 & 0 & 0 & 0 & 0 & 0 & 0 & 0\\  
                 & &  & 0 & 0 & 0 & 0 & 0 & 0 & 0 & 0 & 1\\  
                \end{array}
                \right).
            \]
            Let $x_9 = (1,1,1,\cdots,1,1,1,1)^T$ and $x_{13} = (1,\cdots,1,0,1,1,0,1,1,1,0,1)^T$.  We denote by \(\sigma_7(M')\) the matrix resulting from swapping \(x_9\), \(x_{13}\) and   
            \[
            \varphi_7 = x_9x_9^T + x_{13}x_{13}^T + x_9x_{13}^T + x_{13}x_9^T= 
            \left(
                \begin{array}{*{12}{c}}
                0 & \cdots & 0\\ 
                \vdots & \ddots &\vdots\\ 
                0&\cdots &0  \\
                 & &  & 1 & 0 & 0 & 1 & 0 & 0 & 0 & 1 & 0\\  
                 & &  & 0 & 0 & 0 & 0 & 0 & 0 & 0 & 0 & 0\\  
                 & &  & 0 & 0 & 0 & 0 & 0 & 0 & 0 & 0 & 0\\  
                 & &  & 1 & 0 & 0 & 1 & 0 & 0 & 0 & 1 & 0\\  
                 & &  & 0 & 0 & 0 & 0 & 0 & 0 & 0 & 0 & 0\\  
                 & &  & 0 & 0 & 0 & 0 & 0 & 0 & 0 & 0 & 0\\  
                 & &  & 0 & 0 & 0 & 0 & 0 & 0 & 0 & 0 & 0\\  
                 & &  & 1 & 0 & 0 & 1 & 0 & 0 & 0 & 1 & 0\\  
                 & &  & 0 & 0 & 0 & 0 & 0 & 0 & 0 & 0 & 0\\  
                \end{array}
                \right), 
            \]
            and we cyclically shift row of $\sigma_7(M')$ to obtain the matrix $N'^T$:
            \[
            N'^T =P_2\sigma_7(M') =\begin{pmatrix}
            \sigma_7(M')_2 \\
            \sigma_7(M')_3 \\
            \vdots \\
            \sigma_7(M')_k \\
            \sigma_7(M')_1
            \end{pmatrix}, 
            ~
            \mbox{ where~} 
            P_2 = \begin{pmatrix}
            0 & 1 & 0 &\cdots&0\\
            0 & 0 & 1 &\cdots&0\\
            \vdots &\vdots &\vdots&\ddots&\vdots \\
            0 & 0 & 0 &\cdots&1\\
            1 & 0 & 0 &\cdots&0
            \end{pmatrix}.
            \]
            Then we have 
            \[
            M'N' = 
            \left(
            \begin{array}{*{13}{c}}
   0 & 0 & 0 & 0 & \cdots& \cdots & \cdots & \cdots & \cdots & \cdots & \cdots & 0& 1 \\
   1 & \cdots & 0 & 0&\cdots & \cdots & \cdots & \cdots& \cdots & \cdots & \cdots& 0 & 0 \\
   \vdots & \ddots & \ddots & \vdots& \cdots & \cdots & \cdots & \cdots &\cdots& \cdots& \cdots & \vdots & \vdots \\
   0 & \cdots & 1 & 0 & \cdots & \cdots & \cdots & \cdots & \cdots & \cdots& \cdots& 0 & 0 \\
    & &  & 1 & 0 & 0 & 0 & 0 & 0 & 1 & 1 & 0 & 0 \\
    & &  & 0 & 0 & 0 & 0 & 0 & 0 & 0 & 0 & 0 & 0 \\
    & &  & 0 & 0 & 0 & 0 & 0 & 1 & 0 & 0 & 0 & 0 \\
    & &  & 0 & 0 & 0 & 1 & 0 & 0 & 1 & 1 & 0 & 0 \\
    & &  & 0 & 0 & 0 & 0 & 1 & 0 & 0 & 0 & 0 & 0 \\
    & &  & 0 & 0 & 1 & 0 & 0 & 0 & 0 & 0 & 0 & 0 \\
    & &  & 1 & 0 & 0 & 1 & 0 & 0 & 0 & 0 & 0 & 0 \\
    & &  & 1 & 0 & 0 & 1 & 0 & 0 & 0 & 1 & 0 & 0 \\
    & &  & 0 & 0 & 0 & 0 & 0 & 0 & 0 & 0 & 1 & 0
            \end{array}
            \right)
            \hspace{-0.4em}
            \raisebox{7em}{$\left.\rule{0pt}{2.5em}\right\}$}
            \hspace{0.3em}
            \raisebox{7em}{$I_{k-10},$}
            \]

            \[
            I_k + M'N' = 
            \left(
            \begin{array}{*{13}{c}}
   1 & 0 & 0 & 0 & 0& \cdots & \cdots & \cdots & \cdots & \cdots & \cdots & 0& 1 \\
   1 & 1 & 0 & 0&0& \cdots & \cdots & \cdots& \cdots & \cdots & \cdots& 0 & 0 \\
   \vdots & \ddots & \ddots & \vdots& \vdots & \cdots & \cdots & \cdots &\cdots& \cdots& \cdots & \vdots & \vdots \\
   0 & \cdots & 1 & 1 & 0 & \cdots & \cdots & \cdots & \cdots & \cdots& \cdots& 0 & 0 \\
   0 &\cdots & 0 & 1 & 1 & 0 & 0 & 0 & 0 & 1 & 1 & 0 & 0 \\
   0 &\cdots & 0 & 0 & 0 & 1 & 0 & 0 & 0 & 0 & 0 & 0 & 0 \\
   0 &\cdots & 0  & 0 & 0 & 0 & 1 & 0 & 1 & 0 & 0 & 0 & 0 \\
   0 &\cdots & 0  & 0 & 0 & 0 & 1 & 1 & 0 & 1 & 1 & 0 & 0 \\
   0 &\cdots & 0  & 0 & 0 & 0 & 0 & 1 & 1 & 0 & 0 & 0 & 0 \\
   0 &\cdots & 0 & 0 & 0 & 1 & 0 & 0 & 0 & 1 & 0 & 0 & 0 \\
   0 &\cdots & 0  & 1 & 0 & 0 & 1 & 0 & 0 & 0 & 1 & 0 & 0 \\
   0 &\cdots & 0  & 1 & 0 & 0 & 1 & 0 & 0 & 0 & 1 & 1 & 0 \\
   0 &\cdots & 0 & 0 & 0 & 0 & 0 & 0 & 0 & 0 & 0 & 1 & 1
            \end{array}
            \right)
            \]

            \end{enumerate}    
            
}        
        
        \end{enumerate}
    \end{enumerate}
\end{enumerate}

        Now, let $\D$ be the code generated by $H = (N' : I_{n-k})$. Since $I_k + M'N'$ is invertible, $(\C, \D)$ forms an LCP. Moreover, since $N'^T$ is obtained from $M$ through column permutations and row shifts, $\D^\perp$ has the same parameters as $\C$, meaning both are optimal codes with minimum distance $d$. Therefore, the security parameter $\min\{d(\C), d(\D^\perp)\} = d$ achieves the theoretical maximum $d_L(n,k)$, proving the optimality of the LCP $(\C, \D)$. 

        This completes the proof.
\end{proof}

\begin{example}\label{ex:LCPs-construction}
Let $k = 7$, $d = 49$, $g_2(k, d) = 101$. Note that $h=1$ and hence we can delete a 4-dimensional subspace $U_1$  and its base is listed as follows: 
$\mathcal{B} \setminus \{ \left( {0,1,0,\ldots ,0,1}\right) \ \cup \left( {0,0,1,\ldots ,0,1}\right) \} $.

Choose a basis for $W$ as 
$w_1 = (1, 0, 0, 0, 0, 0, 0)^T, w_2=(0, 1, 0, 0, 0, 0, 0)^T, w_3=(0, 0, 1, 0, 0, 1, 1)^T,w_4=(0, 0, 0, 1, 1, 0, 0)^T$. The number of columns deleted  is $|F_1|=|W \setminus A| = 11$, then $F_1=W \setminus A$ can be shown below:
\[
F_1= 
\left(
\begin{array}{*{11}{c}}
0& 0& 0& 0& 1& 1& 1& 1& 1& 1 &1\\
0& 1& 1& 1& 0& 0& 0& 1& 1& 1 &1\\
1& 0& 1& 1& 0& 1& 1& 0& 0& 1 &1\\
1& 1& 0& 1& 1& 0& 1& 0& 1& 0 &1\\
1& 1& 0& 1& 1& 0& 1& 0& 1& 0 &1\\
1& 0& 1& 1& 0& 1& 1& 0& 0& 1 &1\\
1& 0& 1& 1& 0& 1& 1& 0& 0& 1 &1 
\end{array}
\right).
\]
In this case we can obtain a Belov code with parameters $[101, 7, 49]$.  Then we have
\[
MM^T=\begin{pmatrix}
 0& 0& 0& 0& 0& 0& 0  \\ 
 0& 0& 0& 0& 0& 0& 0  \\
 0& 0& 0& 0& 0& 1& 1  \\
 0& 0& 0& 0& 1& 0& 0  \\ 
 0& 0& 0& 1& 0& 0& 0  \\ 
 0& 0& 1& 0& 0& 0& 1  \\
 0& 0& 1& 0& 0& 1& 0
\end{pmatrix},
\]
we swap columns $x_3 = (1,1,0,0,0,1,0 )^T$ and $x_4 = (1,1,1,0,0,0,1)^T$ and denote by \(\sigma_2(M')\) the matrix resulting from swapping \(x_3\) and \(x_4\). The matrix $N'^T$ is constructed as follows:
\[
N'^T = P_1\sigma_2(M') =  \begin{pmatrix}
\sigma_2(M')_3 \\
\sigma_2(M')_4 \\
\vdots \\
\sigma_2(M')_k \\
\sigma_2(M')_1  \\
\sigma_2(M')_2
\end{pmatrix},
~
\mbox{ where~} 
P_1 = \begin{pmatrix}
0 & 0 & 1 &\cdots&0\\
\vdots &\vdots &\vdots&\ddots&\vdots \\
0 & 0 & 0 &\cdots&1\\
1 & 0 & 0 &\cdots&0\\
0 & 1 & 0 &\cdots&0
\end{pmatrix}.
\]
The resulting matrix $M'N'$ is strictly lower triangular:
\[
M'N'= \begin{pmatrix} 
0\\ 
0&0\\
1&0&0\\
0&0&1&0\\ 
0&1&0&0&0\\ 
0&0&0&1&0&0 \\
0&0&0&0&1&0&0
\end{pmatrix}.
\]
It is easy to see the matrix $I_k + M'N'$ must be invertible. Since both $\C$ and $\D^\perp$ have parameters $[101, 7, 49]$, the security parameter is:
$
d_{LCP} = \max \{ \min\{d(\C), d(\D^\perp)\} \} = 49.
$
\end{example}

\begin{example}\label{ex:LCPs-construction}
Let $k = 5$, $d = 18$, $g_2(k, d) = 37$. Note that $h=2$ and hence we can delete a 4-dimensional subspace $U_1$  and its base is listed as follows: 
$\mathcal{B} $.

Choose a basis for $W$ as 
$w_1 = (1, 0, 0, 0, 0)^T, w_2=(0, 1, 0, 0, 0)^T, w_3=(0, 0, 1, 0, 0)^T,w_4=(0, 0, 0, 1, 0)^T$. The number of columns deleted  is $|F_1|=|W \setminus A| = 10$, then $F_1=W \setminus A$ can be shown below:
\[
F_1= 
\begin{pmatrix}
 0& 0& 0& 0& 1& 1& 1 &1 &1 &1 \\ 
 0& 1& 1& 1& 0& 0& 0 &1 &1 &1\\
 1& 0& 1& 1& 0& 1& 1 &0 &0 &1\\
 1& 1& 0& 1& 1& 0& 1 &0 &1 &0\\ 
 0& 0& 0& 0& 0& 0& 0 &0 &0 &0 
\end{pmatrix}
\]
In this case we can obtain a Belov code with parameters $[37, 5, 18]$, its generator matrix is
\[
G_{Belov}=
\setlength{\arraycolsep}{1.8pt}
\left(
\begin{array}{*{37}{c}}
    0& 0& 0& 0& 0& 0& 0& 0& 1& 1& 1& 1& 1& 1& 1& 1& 0& 0& 0& 0& 0& 0& 0& 0& 0& 0& 0& 1& 1& 1& 1& 1& 1& 1& 1& 1& 1\\
    0& 0& 0& 0& 1& 1& 1& 1& 0& 0& 0& 0& 1& 1& 1& 1& 0& 0& 0& 0& 0& 0& 1& 1& 1& 1& 1& 0& 0& 0& 0& 0& 1& 1& 1& 1& 1\\
    0& 0& 1& 1& 0& 0& 1& 1& 0& 0& 1& 1& 0& 0& 1& 1& 0& 0& 0& 1& 1& 1& 0& 0& 0& 1& 1& 0& 0& 0& 1& 1& 0& 0& 1& 1& 1\\
    0& 1& 0& 1& 0& 1& 0& 1& 0& 1& 0& 1& 0& 1& 0& 1& 0& 1& 1& 0& 0& 1& 0& 0& 1& 0& 1& 0& 0& 1& 0& 1& 0& 1& 0& 1& 1\\
    1& 0& 0& 1& 0& 1& 1& 0& 0& 1& 1& 0& 1& 0& 0& 1& 1& 0& 1& 0& 1& 1& 0& 1& 1& 1& 1& 0& 1& 1& 1& 1& 1& 1& 1& 0& 1\\
\end{array}
\right).
\]

Then we have
\[
MM^T=\begin{pmatrix}
 1& 1& 1& 1& 0  \\ 
 1& 1& 1& 1& 0  \\
 1& 1& 1& 1& 0  \\
 1& 1& 1& 1& 0  \\ 
 0& 0& 0& 0& 1  \\ 
\end{pmatrix},
\]
we swap columns $x_5 = (1,1,1,1,1)^T$ and $x_6 = (0,0,0,0,1)^T$ and denote by \(\sigma_3(M')\) the matrix resulting from swapping \(x_5\) and \(x_6\). The matrix $N'^T$ is constructed as follows:
\[
N'^T = P_2\sigma_3(M') =  \begin{pmatrix}
\sigma_3(M')_2 \\
\sigma_3(M')_3 \\
\vdots \\
\sigma_3(M')_k \\
\sigma_3(M')_1  
\end{pmatrix},
~
\mbox{ where~} 
P_2 = \begin{pmatrix}
0 & 1 & 0 &\cdots&0\\
0 & 0 & 1&\cdots&0\\
\vdots &\vdots &\vdots&\ddots&\vdots \\
0 & 0 & 0 &\cdots&1\\
1 & 0 & 0 &\cdots&0
\end{pmatrix}.
\]
The resulting matrix $M'N'$ is strictly lower triangular:
\[
M'N'= \begin{pmatrix} 
0\\ 
0&0\\
0&0&0\\
0&0&0&0\\ 
0&0&0&1&0
\end{pmatrix}.
\]
It is easy to see the matrix $I_k + M'N'$ must be invertible. Since both $\C$ and $\D^\perp$ have parameters $[37, 5, 18]$, the security parameter is:
$
d_{LCP} = \max \{ \min\{d(\C), d(\D^\perp)\} \} = 18.
$
\end{example}

As shown in Table \ref{table:parameters-comparison}, our Belov-based approach significantly reduces the minimum distance lower bound compared with the Solomon-Stiffler code construction. For example, when $k=6$, the Solomon-Stiffler code requires a minimum distance $d \ge 96$ and $n \ge 189$. In contrast, our Belov code-based construction with $u_1 \ge 4, \theta = 4$ only requires $d \ge 41$ and $n \ge 84$, demonstrating a substantial improvement over the existing results.

\begin{table}[!htbp] 
\begin{center} 
\caption{Minimum Distance Lower Bounds of Solomon-Stiffler and Belov Codes}
\label{table:parameters-comparison}
\begin{tabu}{|c|c|c|c|c|c|}
\hline
\multirow{2}{*}{} & Solomon-Stiffler Code & \multicolumn{2}{c|}{Belov Code} \\
\hline
&$d \geq \lceil (k-1)/2 \rceil 2^{k-1}, h\ge2$ & $u_1 \ge 3, \theta = 3$  & $u_1 \ge 4, \theta = 4$\\
\hline
$k=5$& $d \geq 32, n \geq 62$ & $d \ge21, n \ge 43$ & $h \ge2, d \ge17, n \ge 36$ \\
\hline
$k=6$& $d \geq 96, n \geq 189$ & $d\ge45, n\ge 91$ & $h \ge2,d \ge41, n \ge 84$\\
\hline
$k=7$& $d \geq 192, n \geq 381$  & $d\ge53, n\ge 108$ & $h \ge1, d \ge49, n \ge 101$\\
\hline
$k=8$& $d \geq 512, n \geq 1020$ & $d\ge109, n\ge 220$ & $h \ge 1 , d \ge105, n \ge 213$\\
\hline
$k = 9$ & $d \geq 1024, n \geq 2044$ & $d \geq 221, n \geq444 $ &  $h \ge 1 , d \ge217, n \ge 437$\\
\hline
$k = 10$  &$d \geq 2560, n \geq 5115$ & $d \geq 445, n \geq 892$ & $h \ge 1, d\ge 441, n \ge 892$\\
\hline
\end{tabu}
\end{center} 
\end{table}  

The results clearly show that Belov codes with $\theta=4$ enables the generation of optimal binary LCPs with smaller minimum distances,  while still meeting the Griesmer bound. Compared with the Solomon-Stiffler code construction and the $\theta=3$ Belov code construction, our scheme achieves a smaller parameter range.
These results enrich the family of optimal binary LCPs and provide a more efficient and practical coding scheme for applications in secure communication and side-channel attack resistance.

\end{document}